\documentclass[prl,floatfix,twocolumn,showpacs,amsmath,amssymb]{revtex4}
\usepackage{graphicx}
\usepackage{dcolumn}
\usepackage{bm}
\usepackage{epsfig,psfrag,amsmath,amssymb,float}
\input{epsf}
\begin{document}

\title{Superconducting fluctuations in the Luther-Emery liquid}

\author{Edmond~Orignac}

\affiliation{Laboratoire de Physique Th\'eorique de l'\'Ecole
Normale Sup\'erieure CNRS-UMR8549, F-75231 Paris Cedex 05, France}

\author{Didier~Poilblanc}

\affiliation{Laboratoire de Physique Th\'eorique CNRS-FRE2603,
Universit\'e Paul Sabatier, F-31062 Toulouse, France}

\date{\today}

\begin{abstract}

The single-particle superconducting Green's functions of a
Luther-Emery liquid is computed  by bosonization techniques. Using
a formulation introduced by Poilblanc and Scalapino [{\sl
Phys.~Rev.~B} {\bf 66}, 052513 (2002)], an asymptotic expression
of the superconducting gap is deduced in the long wavelength and
small frequency limit. Due to superconducting phase fluctuations,
the gap exhibits as a function of size $L$ a $(1/L)^{1/2K_\rho}$
power-law decay as well as an interesting singularity at the
spectral gap energy. Similarities and differences with the 2-leg
t-J ladder are outlined.

\end{abstract}

\maketitle

\begin{figure}
\begin{center}
\epsfig{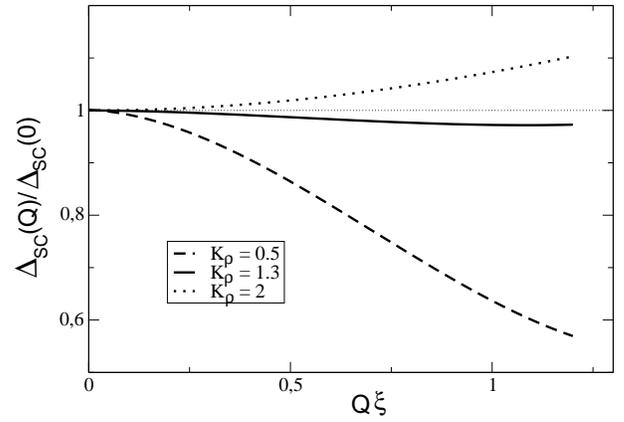}
\caption{Superconducting gap (normalized to its $Q=0$ value) vs
$Q\xi$. } \label{fig:gap1}
\end{center}
\end{figure}

Superconductivity in cuprate materials has sparked an important
theoretical effort to investigate superconducting order in
strongly correlated materials. A long-ranged superconducting
ground state (GS) is characterized by a non-zero superconducting
Gorkov's off-diagonal one-electron Matsubara Green
function\cite{schrieffer64_ch7} defined as
\begin{equation}
F({\bf q},\tau)=-\langle T {c}_{{\bf -q},-\sigma}(\tau/2)
{c}_{{\bf q},\sigma}(-\tau/2)\rangle\, . \label{eq:fk_of_tau}
\end{equation}
As proposed recently~\cite{poilblanc_gap_supra}, basing on the Nambu-Eliashberg
theory\cite{schrieffer64_ch7}, the Matsubara
frequency-dependent gap function $\Delta_{\rm SC}({\bf q},\omega_n)$
can be directly computed from the diagonal and off-diagonal
Green functions  $G({\bf q}, \omega_n)$ and $F({\bf q},\omega_n)$ as
\begin{equation}
\Delta_{\rm SC}({\bf q},\omega_n)= \frac{2i\omega_n F({\bf
q},\omega_n)}{G({\bf q},\omega_n)-G({\bf q},-\omega_n)} \,
.\label{eq:delta}
\end{equation}
Such an expression has been used in numerical computations for the
two-dimensional t-J model\cite{poilblanc_gap_supra} and for
two-leg t-J ladders\cite{poilblanc_gap_2leg}. A one dimensional
system such as the two-leg t-J ladder is actually far remote from
a normal Fermi liquid or a conventional
superconductor\cite{dagotto_2ch_review,schulz_moriond,dagotto_supra_ladder_review}.
The t-J ladder belongs to the the class of Luther-Emery (LE)
liquids\cite{schulz_houches_revue}, which are characterized by
spin/total charge separation, with a spectral gap in spin
excitation and gapless charge
excitations\cite{numerics_tJ_ladder}. The gapless charge
excitations lead to a perfect metallic
conductivity\cite{mikeska_supra_1d} without any long range
superconducting order\cite{mermin_wagner_theorem}. The spectral
gap in the spin excitations generates a gap in the one-electron
spectral functions\cite{voit_le_spectral,tsvelik_spectral_cdw}. In
an infinite system, the absence of long range superconducting
order implies that the correlation function (\ref{eq:fk_of_tau})
has to vanish. However, in a finite system, the function
(\ref{eq:fk_of_tau}) does not vanish and the quantity defined in
(\ref{eq:delta}) may provide useful information on the pairing
processes\cite{poilblanc_gap_2leg}.  As a complementary study to
numerical studies,  it is interesting to investigate analytically
the behavior of $\Delta_{SC}({\bf q},\omega)$ in a finite size LE
liquid. A complication in the case of the two-leg ladder is that
interband (but not total) charge excitations contribute to the
formation of the spin-gap\cite{schulz_moriond}. A simpler example
of a LE liquid is afforded by the $U<0$ Hubbard
chain\cite{luther_exact,andrei_trieste93} or the related spin-gap
phase of the $t-J$ chain\cite{nakamura_tJ} in which the gap is
developed purely in the spin mode. Moreover, this model of a LE
liquid is integrable both on the lattice \cite{andrei_trieste93}
and in the continuum limit \cite{andrei_oNmodel}. Progress in the
theory of integrable systems has made it possible to calculate
correlation functions using the form factor
expansion\cite{karowski_ff}. In the case of gapful systems, it has
been shown that the first few terms of the expansion usually lead
to good approximations of the physical
quantities\cite{controzzi_mott}. In the present paper, we will
calculate $\Delta_{\rm SC}$ in the case of an attractive
one-dimensional Hubbard chain. The Hamiltonian reads:
\begin{eqnarray}
  \label{eq:lattice}
  H=-t\sum_{p=1 \atop\sigma}^N(c^\dagger_{p+1,\sigma}c_{p,\sigma}+\text{H. c.})
  + U \sum_{p=1}^N n_{p,\uparrow} n_{p,\downarrow}.
\end{eqnarray}
Although the Hubbard model in one-dimension is integrable, its form
factors have proved very difficult to
obtain\cite{essler_ff_hubbard}. Since we are mostly interested in
the low-energy and long-wavelength, it  is  more convenient to take the
continuum limit. The continuum
Hamiltonian $H=H_\rho+H_\sigma$ reads:
\begin{eqnarray}
\label{eq:bosonisation} &H_\rho&=v \int_0^L \frac{dx}{2\pi} [K_\rho
(\pi\Pi_\rho)^2
+\frac{1}{K_\rho}(\partial\phi_\rho)^2] \, ,  \\
&H_\sigma&= \int_0^L \frac{dx}{2\pi} [v \{(\pi\Pi_\sigma)^2
+(\partial\phi_\sigma)^2\}-\frac{U}{\pi\alpha}\cos{\sqrt{8}\phi_\sigma}]
\, , \nonumber
\end{eqnarray}
\noindent where $\phi_\rho$ and $\phi_\sigma$ are the
 bosonic fields describing  respectively the spin and charge
 excitations, $\Pi_\rho$ and $\Pi_\sigma$ their
conjugate variables, $\alpha$ is a short distance cut-off ($\sim$
lattice spacing) and $L\sim N\alpha$ is the system size. Note that
we assume here a unique velocity $v$ for spin and charge
excitations. The spin excitations are described by the sine-Gordon
model. The term $\cos \sqrt{8}\phi_\sigma$ is marginally relevant
and opens a spin gap ${\cal M}$. The excitations above the ground
state are known \cite{andrei_oNmodel} to consist only of massive
solitons (spinons) carrying the spin $\pm 1/2$ and charge zero, in
agreement with the exact solution of the lattice Hubbard model
\cite{andrei_trieste93}. It is important to note that with the
usual definition of the spin gap $\Delta_\sigma$ as the gap
between the ground state and the lowest
 triplet excited state, $\Delta_\sigma=2{\cal M}$.
The fermion operators can be decomposed as
$c_{p,\sigma}=\sqrt{\alpha}\sum_r e^{ik_F r x} \psi_{r,\sigma}(x)$
($x=p\alpha$, $r=\pm$ or $\pm 1$) and the fermionic fields
$\psi_{r,\sigma}$ can be expressed as \cite{schulz_houches_revue}:
\begin{eqnarray}
  \label{eq:fermion-field}
  \psi_{r,\sigma}(x)=\frac{e^{\frac{i}{\sqrt{2}}(\theta_\rho-r\phi_\rho)}}{\sqrt{2\pi\alpha}}
  e^{\frac{i}{\sqrt{2}}\sigma(\theta_\sigma-r\phi_\sigma)} \, ,
\end{eqnarray}
where $\theta_\rho$ and $\theta_\sigma$ are the dual fields.
Fermion propagators thus factorize into the product of a holon
propagator and a spinon propagator. Since the charge Hamiltonian
is quadratic, the calculation if the holon propagator is
straightforward. For the spinon propagator, one needs to know the
expression of the  form factors of soliton-creating operators
$e^{\pm i (\theta_\sigma-r\phi_\sigma)}$ in the sine-Gordon model.
These have been obtained recently in infinite
volume\cite{lukyanov_soliton_ff} and these results have been
applied to the calculation of the diagonal Green's
functions\cite{tsvelik_spectral_cdw} confirming the ansatz of
Ref.\cite{voit_le_spectral}. In our case, we need form factors in
a system of finite size. However, if the system size $L$ is much
larger than the correlation length $\xi=v/{\cal M}$, since the
dominant contribution to $\Delta_{SC}(q,\omega)$ comes from
integration on the region $x,v\tau\ll \xi$, it is a good
approximation to replace finite volume form factors by infinite
volume ones. Keeping this in mind, we define the diagonal and
off-diagonal Green's functions as:
\begin{eqnarray}
  \label{eq:greendef}
  G(x,\tau)=-\langle T_\tau \psi_{r,\sigma}(x,\tau)
  \psi^\dagger_{r,\sigma}(0,0)\rangle, \\
  F(x,\tau)=-\langle T_\tau \psi_{r,\sigma}(x,\tau)
  \psi_{-r,-\sigma}(0,0)\rangle.
\end{eqnarray}
Following \cite{tsvelik_spectral_cdw},
the asymptotic (long-distance) behaviors of
the diagonal and off-diagonal Green's functions is then obtained
as,
\begin{eqnarray}
\label{eq:green} &G(x,\tau)&=\frac{Z_1}{2\pi^2} (\frac{\pi
v}{2\alpha{\cal M}})^\frac{1}{2} \frac{1}{ix-v\tau}
(\frac{\alpha}{\rho})^{2\eta}
e^{-{\cal M}\rho/v}\, ,  \\
&F(x,\tau)&\simeq -\frac{Z_1}{2\pi^2\alpha}
(\frac{2\pi\alpha}{L})^\frac{1}{2K_\rho} (\frac{\pi v}{2\rho{\cal
M}})^\frac{1}{2} (\frac{\alpha}{\rho})^{2\eta'} e^{-{\cal
M}\rho/v}   \, , \nonumber
\end{eqnarray}
\noindent where $\rho=\sqrt{x^2+(v\tau)^2}$,
$\eta=\frac{1}{8}(K_\rho+1/K_\rho-2)$,
$\eta'=\frac{1}{8}(K_\rho-1/K_\rho)$, $Z_1$ is a dimensionless
constant calculated in \cite{lukyanov_soliton_ff} and ${\cal M}$
the single particle gap.  These expressions show that $G$ and $F$
have a similar exponential decay at large distances above the
characteristic length-scale $\xi=v/{\cal M}$. More rapidly
decaying terms like $e^{-3{\cal M}\rho/v}$ and higher that involve
more than one spinon in the intermediate state have been
neglected. Note that the SC Green's function decays with system
size like as $(1/L)^\frac{1}{2K_\rho}$ due to SC phase
fluctuations. However, apart from this overall {\it power-law}
scaling factor, we expect the SC gap $\Delta_{\rm SC}$ to remain
finite and bear interesting $q$ and $\omega$ dependence. Using the
Fourier transforms of the above expressions~(\ref{eq:green}), the
momentum and Matsubara-frequency dependence of the gap is given by
\begin{equation}
\label{eq:gap} \Delta_{\rm SC} (q,\omega_n)= {\cal C} {\cal M}
(\frac{2\pi\xi}{L})^\frac{1}{2K_\rho} \frac{{}_2F_1(\frac{3}{4}
-\eta',\frac{5}{4}-\eta';1;y)}{{}_2F_1(1-\eta,\frac{3}{2}
      -\eta;2;y)}\, ,
\end{equation}
where $y=-(Q\xi)^2$, $Q^2=(q-k_F)^2+(\omega_n)^2/v^2$, ${}_2F_1$
is the hypergeometric function\cite{abramowitz_math_functions} and ${\cal
C}=\Gamma(\frac{3}{2}-2\eta') / \Gamma(2-2\eta)$. The approximate expression
(\ref{eq:gap}) although a long-wavelength limit still applies up
to momentum $Q\sim 2\pi/\xi$. Note that $(Q\alpha)^2$ terms are
dropped since $\xi\gg \alpha$. $\Delta_{\rm SC}$ is maximum
(minimum) at $Q=0$ (i.e. at the Fermi momentum) for $K_\rho< 1.57$
($K_\rho> 1.57$) and shows only a very moderate dependence in $Q$
as shown in Fig.~\ref{fig:gap1}. For $Q\rightarrow 0$,
$\Delta_{\rm SC} (q,\omega_n)\simeq \Delta_{\rm SC} (0,0)(1-{\cal
C'}(Q\xi)^2)$.  The constant ${\cal C'}=\frac 1 {64}\left[\frac 1
2
    K_\rho^2 -4 K_\rho -13 + \frac 28 {K_\rho} + \frac 1
    {2K_\rho^2}\right]$ decreases for increasing SC
correlations i.e. for increasing $K_\rho$, e.g.
${\cal C'}\simeq 0.674$, $0.188$ and $0.016$ for $K_\rho=0.5$
(charge density wave regime), $1$ and $1.5$ (superconducting
regime), respectively.

\begin{figure}
\begin{center}
\epsfig{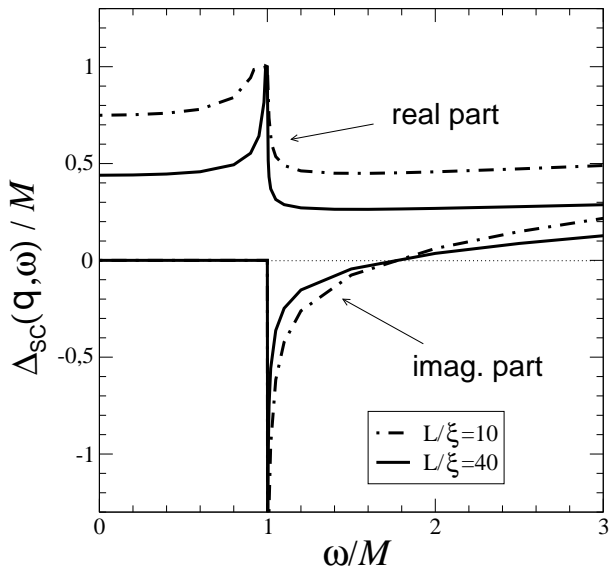} \caption{Real (a)) and
imaginary (b) parts of the superconducting gap vs $\omega$
obtained by bosonization results for a 1D LE liquid (energies are
in units of the spectral (solitonic) gap $\cal M$) for
$K_\rho=1.3$ and $q=k_F$. Data for two lengths are shown.}
\label{fig:super2}
\end{center}
\end{figure}

The real-frequency expression of the gap is obtained by the
analytic continuation $(\omega_n)^2\rightarrow
-(\omega+i\epsilon)^2$ in Eq.~(\ref{eq:gap}). For
$|\omega|>\cal{M}$, the hypergeometric functions have branch cuts,
leading to a nonzero imaginary part in $\Delta_{SC}(0,\omega)$ as
can be seen on Fig.~\ref{fig:super2}(b). Above the threshold, the
gap function presents a singular behavior: $\Delta(0,\omega) \sim
\Delta(0,0) (1-\frac{(\omega+i0)^2}{{\cal M}^2})^{-\frac 1
{2K_{\rho}}}$. The divergence at $\omega={\cal M}$ is in fact
cut-off once $|\omega-{\cal M}|<1/L$ since we are really dealing
with a system of finite size, so that $\Delta_{\rm SC}
(q,\omega\rightarrow{\cal M})={\cal M}$. The full
$\omega$-dependence of $\Delta_{\rm SC}$ plotted in
Fig.~\ref{fig:super2}(a) indeed reveals a strong singularity at
$\omega={\cal M}$ increasingly pronounced as system size is
increased.

In the absence of data for a single attractive Hubbard chain or a
single $t-J$ chain, we compare our results to numerical
calculations on the t-J ladder model at $1/8$-doping and
$J=0.4$\cite{poilblanc_gap_2leg}. The numerical calculations on
$2\times 12$ ladder show that there are two spinon gaps, ${\cal
M}(q_y=0)= 0.08 t$ and ${\cal M}(q_y=\pi)= 0.12 t$ such that
$F(q_y=0,\pi)$, $G(q_y=0,\pi)$ and $\Delta(q_y=0,\pi)$ develop an
imaginary part for $\omega>\Delta(q_y)$. The comparison of these
spinon gaps with the actual spin gap of a $2\times 24$
ladder\cite{CORE} (defined as the gap between the lowest triplet
excitation and the singlet ground state) gives
$\Delta_\sigma(q_y=0)=0.11 t\simeq 2 {\cal M}(q_y=0)$ and
$\Delta_\sigma(q_y=\pi)=0.17 t\simeq  {\cal M}(q_y=\pi)+{\cal
  M}(q_y=0)$. The discrepancies could result from a spinon-spinon
attraction or from the overestimation of the spinon gaps in the $2
\times 12$ ladder. It is also important to note that in
\cite{poilblanc_gap_2leg}, the correlation length is of the order
of magnitude of the system size. We note that no sharp peak is
present in the imaginary part at the threshold, in contrast to the
prediction of (\ref{eq:gap}), but the prediction of a rather
constant behavior of $\Delta$ below the threshold is in agreement
with  (\ref{eq:gap}).

A more detailed comparison between analytic and numerical result
is possible. In the case of the ladder system, away from
half-filling, the gapped modes is expected to present an
approximate $SO(6) \sim SU(4)$ symmetry\cite{schulz_son,
ahn_paik}. This allows a description of the gapped modes by the
SO(6) Gross-Neveu (GN) model\cite{gross_neveu} and a form factor
calculation of the superconducting gap $\Delta({\bf q},\omega_n)$
along the lines of the present paper\cite{ahn_paik}. The novelty
in the case of the $SO(6)$ GN model is that on top of the spinon
excitations of mass ${\cal M}$ (known as kinks in the literature
on the GN model), there are also massive fermion excitations
(bound states) with a mass ${\cal
  M}\sqrt{2}$. This implies a second threshold in $\Delta({\bf
  q},\omega_n)$ at the energy $\omega={\cal M}(1+\sqrt{2})$ besides
the threshold at energy $\omega={\cal M}$. Whether this could be
related to some higher energy features seen in
numerics~\cite{poilblanc_gap_2leg} needs further clarifications.
This will be discussed in more details in a separate publication.

In conclusion, we have computed the fluctuating SC gap of the LE
chain. A simple form is obtained with a factorization into a
power-law factor accounting for SC suppression due to quantum
phase fluctuations multiplied by a function containing the
dynamics of the pairing interaction. We point out some differences
with the case of the 2-leg t-J ladder and suggest that the gapped
sectors of the latter could be better described by the SO(6) GN
model.


\end{document}